\documentclass{article}
\usepackage[dvips]{graphics}
\usepackage{epsfig}
\title{Total, elastic and diffractive cross sections at LHC  \\ in the Miettinen--Pumplin model}
\author{S.~Sapeta\footnote{e-mail: sapeta@th.if.uj.edu.pl}\\
\textit{M. Smoluchowski Institute of Physics,
Jagellonian University, Cracow, Poland}\\ \\
K.Golec-Biernat\footnote{e-mail: golec@ifj.edu.pl}\\
\textit{Institute of Nuclear Physics, Polish Academy of Sciences, Cracow, Poland}\\
\textit{Institute of Physics, University of Rises\'ow, Rzesz\'ow, Poland}
}

\date{}

\begin{document}
\maketitle
Keywords: total cross section, elastic cross section, single diffraction, LHC \\

PACS: 13.60.Hb, 13.85.-t, 13.85.Lg, 13.85.Hd
\begin{abstract}
Predictions for the total, elastic and single diffractive cross sections calculated for the LHC in the framework of the Miettinen-Pumplin model are presented. The total cross section  is expected to be $15 \%$  smaller than that determined by Donnachie and Landshoff in the model with soft pomeron.
The  diffractive cross section is  almost constant in the Tevatron--LHC
energy range.
\end{abstract}

\section{Introduction}

Predictions for the total, elastic and diffractive cross sections at the LHC
energy $14~{\rm TeV}$, one order of magnitude bigger than currently available energy,
are of great importance. Because of that, a number of authors have been trying to
provide such predictions.
The most popular model used for this purpose is the Regge theory  with the soft pomeron  \cite{Donnachie:1992ny}, which describes very well
the existing data  but has one serious drawback -- it violates the unitarity-based Froissart--Martin bound \cite{FMbound}. Nevertheless, because it seems that data from the present experiments are relatively far from this bound, the soft pomeron model is widely accepted.  However, attempts also exist in which cross sections are determined in unitarity preserving descriptions. They are based either on the Regge theory \cite{Covolan, Levin, Ryskin,  Luna:2004gr} or QCD--inspired parameterizations \cite{Block:2000pg}.

In this paper we propose a method of determining the total, elastic and diffractive cross sections at the LHC in the framework of the Miettinen-Pumplin model which preserves
unitarity \cite{Miettinen:1978jb}.  The validity of this model at the Tevatron energy 
$1.8~{\rm TeV}$ has been recently demonstrated \cite{Sapeta:2004av}.

The Miettinen and Pumplin model is based on the Good-Walker picture of soft diffraction \cite{Good:1960ba}.
The state of the incident hadron is expanded into a superposition of states which are eigenstates of diffraction
\begin{equation}
\mid B\rangle\,=\,\sum_{k}C_k\mid\psi_k \rangle \label{expansion}
\end{equation}
\begin{equation}
ImT\mid\psi_k \rangle\,=\,t_k\mid\psi_k \rangle\,,
\end {equation}
where from unitarity: $0\le t_k \le 1$.
If different eigenstates are absorbed by the target with different
intensity, the outgoing state is no longer $\mid B \rangle$ and the inelastic
production of particles takes place. The inelastic diffractive cross section is proportional to the dispersion of the distribution of $t_k$ for the state $\mid B \rangle$.

The basic assumption of Miettinen and Pumplin is that the eigenstates of diffraction are parton states \cite{Miettinen:1978jb}
\begin{equation}
\mid \psi_k \rangle\equiv \mid \vec{b}_1,...,\vec{b}_N,y_1,...,y_N \rangle \label{as}
\end {equation}
where $N$ is the number of partons, and $(y_i,\vec{b}_i)$ are rapidity and impact parameter (relative to the center of the projectile) of parton $i$,
respectively. Therefore, eq.~(\ref{expansion}) takes the form
\begin{equation}
\mid B\rangle\,=\,\sum_{N=0}^{\infty}\int \prod_{i=1}^{N} d^2\vec{b}_i\,dy_i  \,C_N(\vec{b}_1,...,\vec{b}_N,y_1,...,y_N) \mid \vec{b}_1,...,\vec{b}_N,
\ y_1,...,y_N \rangle.
\end{equation}
With a number of assumptions concerning parton distributions and interactions (see \cite{Miettinen:1978jb,Sapeta:2004av}), Miettinen and Pumplin arrived at the following formulae for the differential cross sections
which depend on two parameters only: $\beta[fm^2]$ and $G^2$,
\begin{eqnarray}
\frac{d\sigma_{tot}}{d^2b}\,=\,
2\left(1-\exp\left(-G^2\frac{4}{9}\,
\,e^{-b^2/(3\beta)}\right)\right)
\label{eq:tot}
\end{eqnarray}
\begin{eqnarray}
\frac{d\sigma_{el}}{d^2b}
\,=\,
\left(1-\exp\left(-G^2\,\frac{4}{9}\,e^{-b^2/(3\beta)}\right)\right)^2
\label{eq:el}
\end{eqnarray}
\begin{eqnarray}
\frac{d\sigma_{diff}}{d^2b}
\,=\,
\exp\left(-2\,G^2\,\frac{4}{9}\, e^{-b^2/(3\beta)}\right) \left(\exp\left(G^2\,\frac{1}{4}\,e^{-b^2/(2\beta)}\right)-1\right)\,.
 \label{eq:diff}
\end{eqnarray}
The total cross sections $\sigma_{tot}, \sigma_{el}$ and $\sigma_{diff}$
are obtained after the integration over the impact parameter $b$.
Let us notice that single diffraction is only considered
in the Miettinen--Pumplin model.

The two free parameters $\beta$ and $G^2$ in the presented model
can be calculated for a given energy $\sqrt{s}$,
using the experimental values for $\sigma_{tot}$ and  $\sigma_{el}$. The results of such a
calculation are presented in Table~\ref{beta-GG}. The errors of $\beta$ and $G^2$
are obtained with the help of the total differential method
from the experimental errors of $\sigma_{tot}$ and $\sigma_{el}$. Having determined
the free parameters,
the diffractive cross section can be \textit{predicted} from
eq.~(\ref{eq:diff}). As shown in Table~2, the predictions for $\sigma_{diff}$
are in good agreement with data at the energies of the order of {\rm TeV} \cite{Sapeta:2004av}.

\vskip 0.3cm
\begin{table}[ht]
\begin{center}
\begin{tabular}{|c|c||c|c||c|c|} \hline
Data sets      & $\sqrt{s}~{\rm  [GeV]}$
& $\sigma_{tot}~ {\rm [mb]}$ & $\sigma_{el}~{\rm [mb]}$     &    $\beta~{\rm [fm^2]}$        &        $G^2$  \\ \hline\hline
ISR~\cite{Amos:1985wx} &    30.4     &   $42.13  \pm 0.57$ &  $7.16  \pm 0.34$ & $0.284  \pm 0.018$  &  $2.21  \pm 0.16$ \\ \hline
ISR~\cite{Amos:1985wx} &    52.6     &   $43.32  \pm 0.34$ &  $7.44  \pm 0.32$ & $0.287  \pm 0.016$  &  $2.26  \pm 0.15$ \\ \hline
ISR~\cite{Amos:1985wx}&    62.3     &   $44.12  \pm 0.39$ &  $7.46  \pm 0.32$ & $0.299  \pm 0.016$  &  $2.20  \pm 0.14$ \\ \hline
CDF~\cite{Abe:1993xy} &    546      &  $61.26  \pm 0.93$ & $12.87  \pm 0.30$  & $0.319  \pm 0.015$  &  $3.11  \pm 0.15$ \\ \hline
UA4~\cite{Bozzo:1984rk}&    546       &  $61.90  \pm 1.50$   & $13.30  \pm 0.40$  & $0.313  \pm 0.021$ &  $3.23  \pm 0.22$ \\ \hline
E811~\cite{Avila:1998ej}&   1800       & $71.71  \pm 2.02$  & $15.79  \pm 0.87$ & $0.351  \pm 0.034$  &  $3.39  \pm 0.38$ \\ \hline
CDF~\cite{Abe:1993xy}&     1800       & $80.03  \pm 2.24$   & $19.70  \pm 0.85$ & $0.337  \pm 0.030$  &  $4.20  \pm 0.45$ \\ \hline
\end{tabular}
\end{center}
\vskip -0.2cm
\caption{Parameters of the Miettinen-Pumplin model, $\beta$ and $G^2$, determined from total and elastic $p\bar{p}$
cross sections in the energy range $30-1800~{\rm GeV}$ \cite{Miettinen:1978jb,Sapeta:2004av}.}
\label{beta-GG}
\end{table}
\vskip 0.3cm

With the determined parameters, the scattering matrix in the Miettinen-Pumplin  model,
\begin{equation}
\label{eq:s}
S(b)\,=\,\exp\left(-G^2\frac{4}{9}\,e^{-b^2/(3\beta)}\right)\,,
\end{equation}
is close to the blackness limit ($S(b)\ll 1$) at the central impact parameters at
the Tevatron energy, see Figure~1.
This is also manifest in the saturation of the bound for the differential cross sections \cite{Pumplin:1973cw,Fialkowski:1975ta},
\begin{equation}
\frac{d\sigma_{el}}{d^2b}+\frac{d\sigma_{diff}}{d^2b}\, \le \, \frac{1}{2}  \frac{d\sigma_{tot}}{d^2b}\,,
\end{equation}
taken at the central impact parameter. As seen in
Figure~1, $\sigma_{diff}(0)\ll \sigma_{el}(0)\simeq \sigma_{tot}(0)/2$.
However, the corresponding relation for the integrated cross sections,
\begin{equation}
\label{eq:fmbound}
\sigma_{el}+ \sigma_{diff}\le\sigma_{tot}/2\,,
\end{equation}
is less saturated at Tevatron
(approximately: $16~{\rm mb}+ 9~{\rm mb} < (72/2)~{\rm mb}$)
because the proton is more transparent away from the center.
The peripheral character of diffraction is also clearly demonstrated in Figure~1, since the ratio
$\sigma_{diff}(b)/\sigma_{tot}(b)$ is the biggest for $b$ in the range $1-1.5~{\rm fm}$.

\begin{figure}[t]
\begin{center}
\epsfig{file=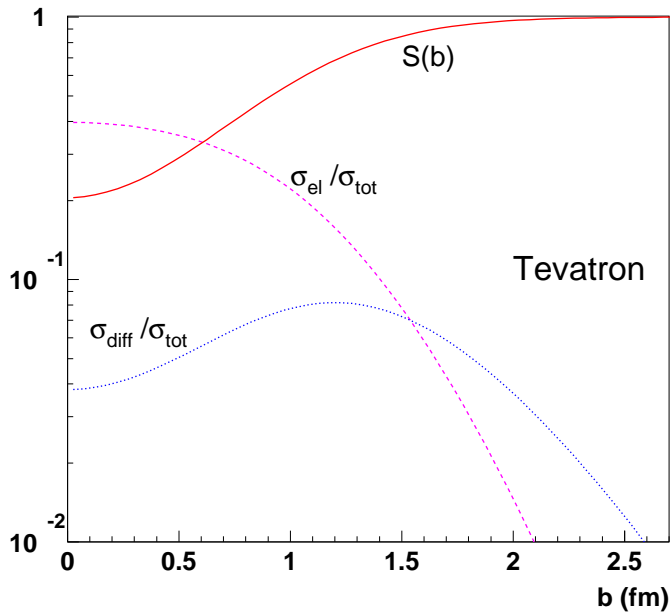, width=10cm}
\end{center}
\vskip -0.5cm
\caption{Impact parameter dependence of the differential cross sections (\ref{eq:tot})-(\ref{eq:diff})
and the scattering matrix (\ref{eq:s}) at the Tevatron energy $1.8$~{\rm TeV}.}
\label{fig:bdep}
\end{figure}

The dependence of the elastic and diffractive cross sections on the momentum transfer $t$ for $|t|\ll 1~{\rm GeV}^2$
is a good test of the Mettinen--Pumplin model. Such a dependence can be calculated
based on  eqs.~(\ref{eq:el}) and (\ref{eq:diff}), after performing Fourier transform
from the impact parameter space into a transverse momentum space.
For the elastic cross section, we find
\begin{eqnarray}
\frac{d\sigma_{el}}{dt}
\,=\, \frac{\pi}{4} \left(\int_0^\infty\!\! db^2\, J_0(b\sqrt{|t|}) \left[1-\exp\left(-G^2\frac{4}{9}\, e^{-b^2/(3\beta)}\right)\right]\right)^2\,,
\end{eqnarray}
while the  single diffractive cross section is given by
\begin{eqnarray}
\nonumber
\frac{d\sigma_{diff}}{dt}
\!\!\!&=&\!\!\! \frac{1}{8}\, \int_0^\infty\!\! db^2 \int_0^\infty\!\! db^{\prime 2}
 \int_0^{2\pi}\!\! d\theta\, J_0(b^{\prime}\sqrt{|t|})\,
\left\{\exp\left(G^2 \frac{1}{4}\,e^{-(b^2
+ \frac{3}{4}b^{\prime 2}+b\, b^\prime \cos\theta)/(2\beta)}\right)-1\right\}
\\\nonumber
\\
&&~~~~~~~~~~~~~
\times\,\exp\left(-G^2\, \frac{4}{9}\, \left(
e^{-b^2/(3\beta)}+
e^{-(b^2+b^{\prime 2}+2 b\,  b^\prime\cos\theta)/(3\beta)}\right)\right).
\end{eqnarray}
Using the parameters from Table~1,
it appears that in the range:
$|t|<0.2~{\rm GeV}^2$, the elastic and diffractive  cross sections can be effectively parametrized by the formula
\begin{equation}
\frac{d\sigma}{dt}\,=\,
\frac{d\sigma}{dt}\Big|_{t=0}\,e^{-B |t|}
\end{equation}
with the slopes $B_{el}$ and $B_{diff}$,
respectively. In Table~\ref{tab:slopes} we show the calculated values of the slope parameters together with the experimental values. As we see, good  agreement is obtained, which undoubtly makes the Miettinen--Pumplin model trustworthy, at least for the values of $|t|$ up to $0.2~{\rm GeV}^2$.

\begin{table}[t]
\begin{center}
\begin{tabular}{|c|c||c|c||c|c||c|c|} \hline
 Data sets &~$\sqrt{s}$~ &~ $\sigma_{diff}$~  &  Experiment
 &~~ $B_{el}$~~ & Experiment &~ $B_{diff}$ ~& Experiment \\ \hline\hline
ISR       & 30.4 & 7.71 &  &12.4  &  $12.70  \pm 0.50$~\cite{Amos:1985wx}   &  8.1  &   \\ \hline
ISR       & 52.6 & 7.82 & & 12.6  &  $13.03  \pm 0.52$~\cite{Amos:1985wx}  &  8.2  &   \\ \hline
ISR       & 62.3 & 8.11 & &13.1  & $13.47  \pm 0.52$~\cite{Amos:1985wx}   &  8.5 & \\ \hline
CDF       & 546  & 8.82 & $7.89  \pm 0.33$~\cite{Abe:1993wu} & 14.8  &  $15.28  \pm 0.58$~\cite{Abe:1993ss} &  9.4  &   \\ \hline
UA4       & 546  & 8.62 &  $9.40  \pm 0.70$~\cite{Bernard:1986yh} & 14.7  &  $15.20  \pm 0.20$~\cite{Bozzo:1984ri}  &  9.2  &   \\ \hline
E811/E710 & 1800 & 9.63 & $9.40  \pm 0.14$~\cite{Amos:1992jw} &16.9  & $16.98  \pm 0.25$~\cite{Amos:1991bp}&  10.4 &
$10.5  \pm 1.8$~\cite{Amos:1992jw} \\ \hline
CDF       & 1800 & 8.87 & $9.46  \pm 0.44$~\cite{Abe:1993wu} & 17.1  &  $16.99  \pm 0.47$~\cite{Abe:1993ss} &  10.2  &   \\ \hline
\end{tabular}
\end{center}
\vskip -0.2cm
\caption{Predictions of the Miettinen--Pumplin model for diffractive cross section  (in ${\rm mb}$), and
elastic and diffractive slopes (in ${\rm GeV^{-2}}$),
together with experimental results at different energies (in GeV) for $p\bar{p}$ collisions.
}
\label{tab:slopes}
\end{table}

\section{Predictions for LHC}

In order to obtain predictions for the total, elastic and single diffractive cross sections at the LHC we have to extrapolate the parameters $\beta$  and $G^2$ to the energy $\sqrt{s}=14~{\rm TeV}$. For this purpose, we plot the dependence of  $\beta$ and $G^2$ on $\ln\sqrt{s}$ for the existing experiments, using the values from Table~\ref{beta-GG}.

\begin{figure}[p]
	\begin{center}
		\rotatebox{270}{\scalebox{0.50}
{\includegraphics{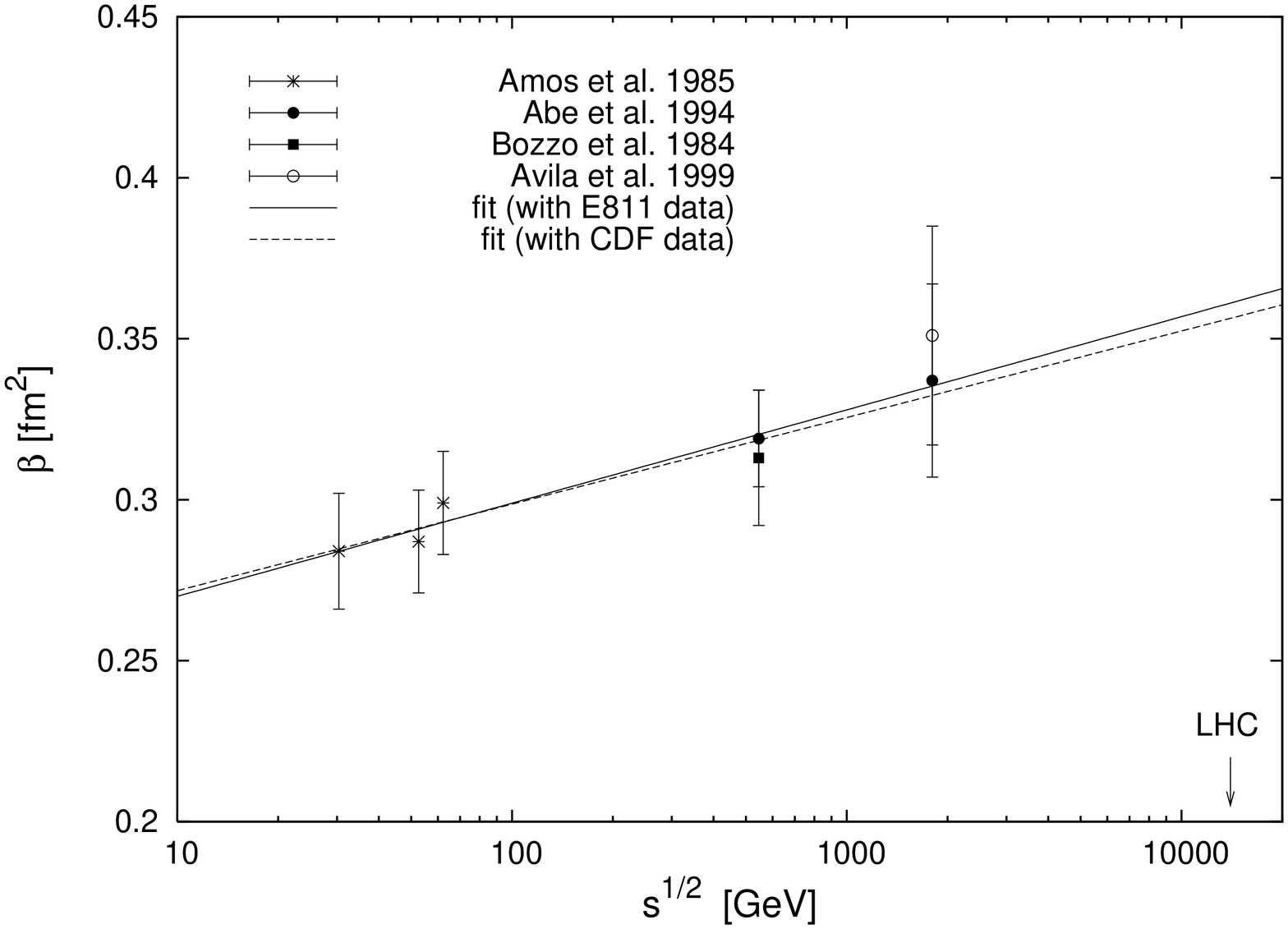}}}
	\end{center}
	\vskip -0.2cm
\caption{Fits of the linear dependence of $\beta$ on $\ln\sqrt{s}$ in the Miettinen-Pumplin model. Data points are from Table~1.}
\label{beta-plot}
\end{figure}

\begin{figure}[hp]
	\begin{center}
		\rotatebox{270}{\scalebox{0.50}
{\includegraphics{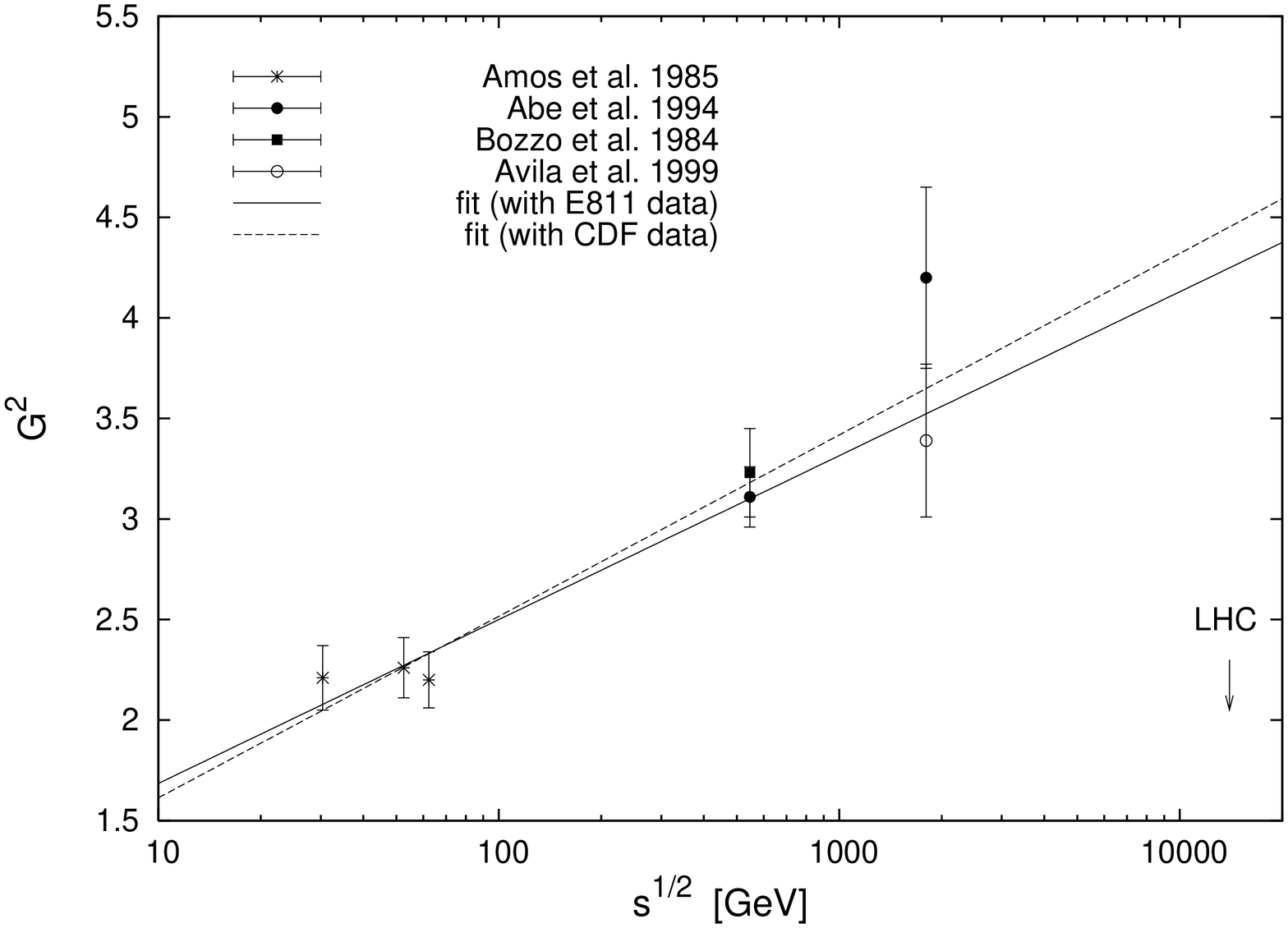}}}
	\end{center}
	\vskip -0.2cm
\caption{Fits of the linear dependence of $G^2$ on $\ln\sqrt{s}$ in the Miettinen-Pumplin model. Data points are from Table~1.}
\label{GG-plot}
\end{figure}

As seen in Figures~ \ref{beta-plot} and \ref{GG-plot}, the parameters depend linearly on $\ln\sqrt{s}$ to a good approximation. Thus, we extrapolate this dependence to the LHC energy by fitting straight lines to the existing data points for $\beta$ and $G^2$ with the errors from Table~1.
Using (\ref{eq:tot}) one can show that with the assumption of the linear dependence, the behaviour of the total cross section is in agreement with the Froissart-Martin bound. Namely, for high $\sqrt{s}$ we obtain
\begin{equation}
\sigma_{tot}\, \propto\, \ln(s)\,\ln(\ln s)\,, \label{froissart}
\end{equation}
which is smaller than $\ln^2 s$.

Fitting the energy dependence,
we faced a well-known problem related to the significant discrepancy between the data for $\sigma_{tot}$ and $\sigma_{el}$ obtained from the E811 \cite{Avila:1998ej} and CDF \cite{Abe:1993xy} experiments at the highest presently available energy $\sqrt{s}=1.8~{\rm TeV}$. Therefore, we decided to consider these two cases separately by performing two fits. The straight lines obtained in this way are shown in Figures~\ref{beta-plot} and \ref{GG-plot}. We see that despite the E811/CDF discrepancy the two alternative fits give similar results.

As an alternative method of the determination of the parameters
$\beta$ and $G^2$, we performed
a four parameter fit to the existing data on $\sigma_{tot}$ and
$\sigma_{el}$, assuming a linear dependence of the parameters on
$\ln\sqrt{s}$, i.e. $\beta=A\ln\sqrt{s}+B$ and $G^2=C\ln\sqrt{s}+D$.
With the E811 data, we found practically the same values of $\beta$ and $G^2$  as before with a very good value of $\chi^2$ of the fit. The fit with the CDF data had worse but still acceptable $\chi^2$.

\begin{table}[t]
\begin{center}
\begin{tabular}{|c||c|c|c||c|c|} \hline
 Scenarios     & $\sigma_{tot}~ {\rm [mb]} $ & $\sigma_{el}~ {\rm [mb]}$ &
 $\sigma_{diff}~ {\rm [mb]}$ & $B_{el}~[{\rm GeV}^{-2}]$ & $B_{diff}~[{\rm GeV}^{-2}]$\\ \hline
with E811 data \cite{Avila:1998ej}  &    $86  \pm 4$    &   $21  \pm 1$   &    $9.5  \pm 0.4$ & 18.8 & 10.6
\\ \hline
with CDF data  \cite{Abe:1993xy}  &   $88  \pm 4$    &   $22  \pm 2$   &    $9.2  \pm 0.5$ & 18.7 & 10.8
\\ \hline
\end{tabular}
\end{center}
\vskip -0.2cm
\caption{Predictions for total, elastic and diffractive cross sections
at the LHC energy $14~{\rm TeV}$, calculated in two scenarios, together with
predictions for elastic and diffractive slopes .}
\label{cs-results}
\end{table}

\begin{figure}[hp]
	\begin{center}
		\rotatebox{270}{\scalebox{0.50}{\includegraphics{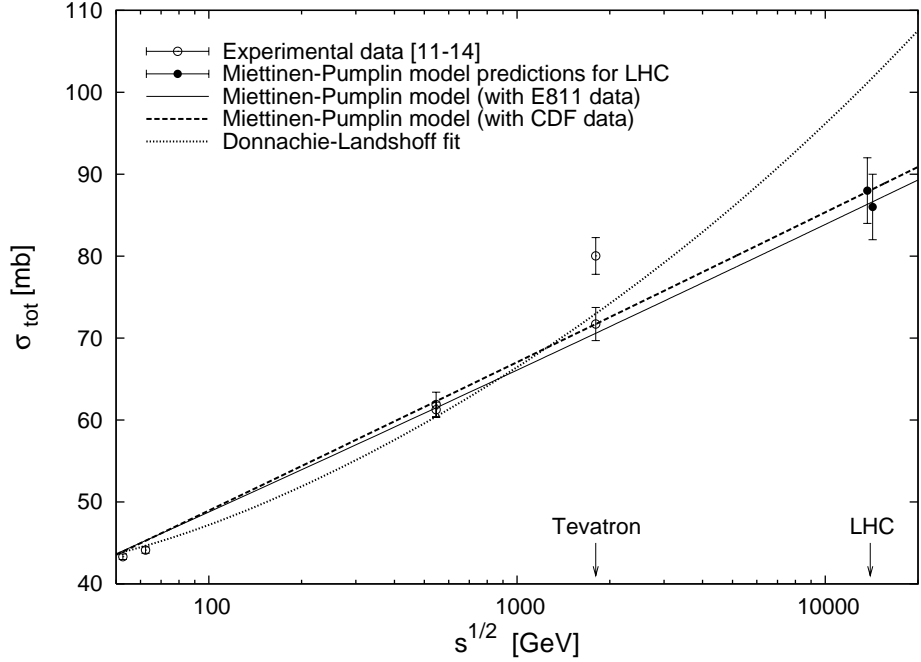}}}
	\end{center}
	\vskip -0.2cm
\caption{Total cross section from the Miettinen-Pumplin model together with the Donnachie-Landshoff prediction. Data points at the LHC energy are predictions  from
Table~2. Experimental data are from \cite{Amos:1985wx, Abe:1993xy, Bozzo:1984rk, Avila:1998ej}.}
 
\label{tot-plot}
\end{figure}

\begin{figure}[hp]
	\begin{center}
		\rotatebox{270}{\scalebox{0.50}{\includegraphics{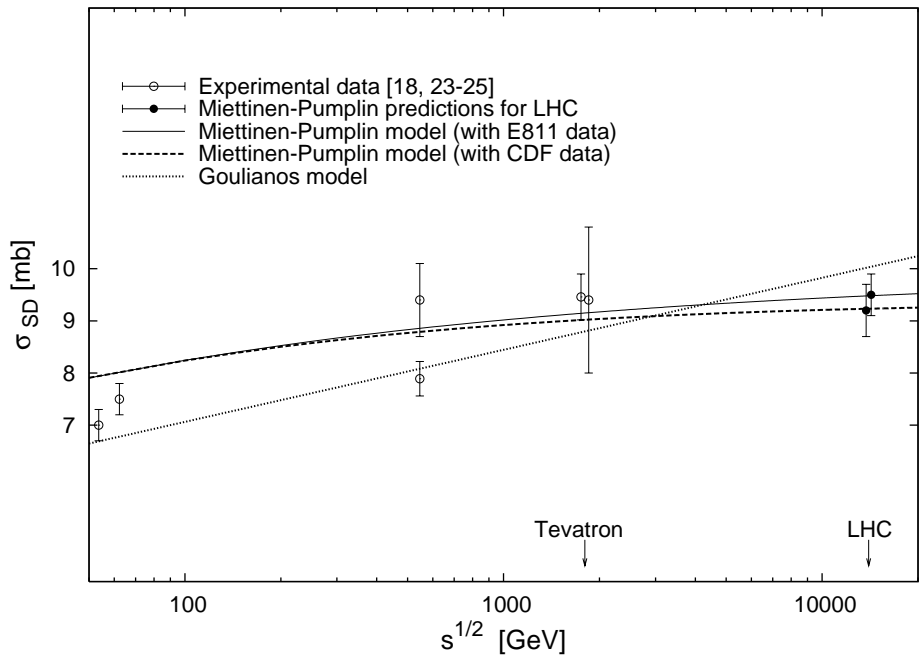}}}
	\end{center}
	\vskip -0.2cm
\caption{Diffractive cross section from the Miettinen-Pumplin model together with  the Goulianos prediction. Data points at the LHC energy are predictions from Table~2. Experimental data are from \cite{Amos:1992jw, Armitage:1982, Bernard:1986yh, Abe:1993wu}.}
\label{diff-plot}
\end{figure}

The predictions for $\sigma_{tot}$, $\sigma_{el}$ and $\sigma_{diff}$ at
the LHC in the the two scenarios are presented in Table~\ref{cs-results}.
The errors
are computed from uncertainties in the determinations of parameters of  the straight lines 
in Figures~\ref{beta-plot} and \ref{GG-plot}, using full covariance matrix.
The dependence of the total and diffractive cross sections on the center-of-mass energy $\sqrt{s}$ in our model (as well as the predicted data points from Table~2) is shown in Figures \ref{tot-plot} and \ref{diff-plot}.
A few comments are in order.

Our prediction for the total cross section at the LHC is $15\%$ smaller than the prediction of Donnachie and Landshoff with the soft pomeron: $101.5~{\rm mb}$  \cite{Donnachie:1992ny}. This difference can be attributed to unitarity constraints present in our approach, and reflected in the behaviour (\ref{froissart}). The result for $\sigma_{tot}$ is also smaller than those given in
\cite{Levin, Ryskin, Luna:2004gr, Block:2001ru} which are above the Donnachie-Landshoff value.
It is also worthwhile to notice that according to our model, the unitarity bound (\ref{eq:fmbound}) is not saturated at the LHC, however the blackness at the central impact parameter is bigger than at Tevatron: $S(b=0) \approx 0.1$.

The diffractive cross section at the LHC is close to that found at Tevatron \cite{Sapeta:2004av}. As seen in Figure~\ref{diff-plot}, $\sigma_{diff}$ is almost constant in the range  starting  from the Tevatron energy. This is a unique prediction of our model which differs from  the prediction of Goulianos \cite{Goulianos:wy}. However, both predictions give a  similar result at the LHC energy. The asymptotic energy behaviour
of the diffractive cross section can be found by inspecting formula (\ref{eq:diff}).
For large values of the parameter $G^2$, $\sigma_{diff}\sim \beta/\sqrt{G^2}$.
Thus with the logarithmic dependence on energy for both parameters, we obtain in the high
energy limit
\begin{equation}
\sigma_{diff}\sim \sqrt{\,\ln s}\,.
\end{equation}
We have also determined the elastic and diffractive slopes at the LHC in the
range $0 < |t| \le 0.2~{\rm GeV}^2$. In both scenarios we obtained practically the same values
which are given in Table~{\ref{cs-results}}.

In summary, we applied the Miettinen-Pumplin model to the determination of the total, elastic and single diffractive cross sections at the LHC. Based on the existing data,
we extracted the energy dependence of two parameters in the model, which was subsequently  extrapolated to the LHC energy. With the  obtained energy dependence, the Froissart-Martin bound is fulfilled. As main results, the prediction for the total cross section at the LHC is
$15\%$ smaller than that  from the soft pomeron model of Donnachie and Landshoff, and
the diffractive cross section is almost constant in the Tevatron--LHC energy range.

\vskip 1cm
\paragraph{Acknowledgements\\ \\}

It is a pleasure to thank Andrzej Bia\l{}as and  Anna Sta\'sto for numerous valuable discussions and critical reading of the manuscript. This research has been supported in part by
the grant of the Polish State Committee For Scientific Research, No.~1~P03B~02828.

\vskip -0.5cm

\end{document}